\documentclass[12pt,preprint]{aastex}

\usepackage{epsfig}
                                                                                
\def\vkm{km s$^{-1}$}

\def\degree{$^\circ$}

\def\arcs#1{$#1''$}
\def\arcsa#1#2{$#1^{\prime\prime}_{^\textrm{.}}#2$}

\def\smassrate{$M_\odot$ yr$^{-1}$}
\def\solarmass{$M_\odot$}

\def\solarlum{$L_\odot$}

\def\Jyb{Jy beam$^{-1}$}
\def\mJyb{mJy beam$^{-1}$}

\def\Jybk{Jy beam$^{-1}$ km s$^{-1}$}

\def\tlabel#1{(\textit{#1})}

\def\cmc{cm$^{-3}$}
\def\cms{cm$^{-2}$}
\def\micron{$\mu$m}
                                                                                
\def\ra#1#2#3#4{#1^\mathrm{h} #2^\mathrm{m} #3^\mathrm{s}_{^\textrm{.}} #4}
\def\dec#1#2#3#4{#1\degr #2\arcmin #3^{\prime\prime}_{^\textrm{.}}#4}

\def\mH{m_\textrm{\scriptsize H}}
\def\Ro{R_\textrm{\scriptsize 0}}

\def\Rin{R_\textrm{\scriptsize in}}
\def\Rout{R_\textrm{\scriptsize out}}
\def\To{T_\textrm{\scriptsize 0}}
\def\no{n_\textrm{\scriptsize 0}}

\def\H2{H$_2$}
\def\N2HP{N$_2$H$^+$}

\def\cCO{C$^{18}$O}

\def\aCO{$^{12}$CO}
\def\bCO{$^{13}$CO}
\def\NH3{NH$_3$}

\def\SOta{$N_J=5_6-4_5$}

\def\putfig#1#2#3{\epsfig{scale=#1,angle=#2,figure=#3}}
\def\putfig#1#2#3{\includegraphics[angle=#1,scale=#2]{#3}}
\def\putfig#1#2#3{}
\def\leftblank#1{}

\begin{document}

\title{A Change of Rotation Profile in the Envelope in the HH 111 Protostellar
System: A Transition to a Disk?}
\author{Chin-Fei Lee\altaffilmark{1}
}
\altaffiltext{1}{Academia Sinica Institute of Astronomy and Astrophysics,
P.O. Box 23-141, Taipei 106, Taiwan; cflee@asiaa.sinica.edu.tw}

\begin{abstract}
The HH 111 protostellar system consists of two Class I sources (VLA 1 and 2) with
putative disks deeply embedded in a flattened envelope at a distance of 400
pc. Here is a follow-up study of this system in
\cCO{} (J=2-1), SO (\SOta), and 1.33 mm continuum at $\sim$ \arcs{1} (400
AU) resolution, and it may show for the first time how a rotationally
supported disk can be formed inside an infalling envelope. The 1.33 mm
continuum emission is seen arisen from both sources, likely tracing the
dusty putative disks around them. In particular, the emission around the VLA
1 source is elongated in the equatorial plane with a radius of $\sim$ 300
AU. The envelope is well seen in
\cCO{}, extending to $\sim$ 7000 AU out from the VLA 1 source, with
the innermost part overlapping with the dusty disk.
It has a differential rotation, with the outer part ($\sim$ 2000--7000 AU)
better described by a rotation that has constant specific angular momentum
and the inner part ($\sim$ 60--2000 AU) by a Keplerian rotation. 
The envelope seems to also have some infall motion that is smaller
than the rotation motion.
Thus, the material in the outer part of the envelope
seems to be slowly spiraling inward with its angular
momentum and the rotation can indeed become
Keplerian in the inner part.
A compact SO emission is seen around the VLA 1 source
with a radius of $\sim$ 400 AU and it may trace a shock
such as an (inner) accretion shock around the disk.
\end{abstract}


\section{Introduction}

Stars are formed inside molecular cloud cores by means of gravitational
collapse. The details of the process, however, are complicated by the
presence of magnetic fields and angular momentum. As a result, in addition to
infall (or collapse), rotation and outflow are also seen toward star-forming
regions. In theory, a rotationally supported disk is expected to form
inside a collapsing core around a protostar, from which part of the material
is accreted by the protostar and part is ejected away. Observationally,
however, when and how a rotationally supported disk is actually formed are
still unclear, because of the lack of detailed kinematic studies inside the
collapsing core.

This paper is a follow-up study of the HH 111 protostellar system. The
properties of this system have been reported in
\citet{Lee2009HH111} and only the important ones are summarized here.
This system is deeply embedded in a
compact molecular cloud core in the L1617 cloud of Orion at a distance of
400 pc. It is a Class I protostellar system with a flattened envelope,
putative disks, and highly collimated jets. At the
center of this system, there are two sources, VLA 1 and VLA 2, with a
projected separation of $\sim$ \arcs{3} (1200 AU) and the
former driving the HH 111 jet \citep{Reipurth1999}.  Previous observation of this system has
shown that the flattened envelope seems to be in transition to a
rotationally supported disk near the VLA 1 source (Lee et al 2009,
hereafter Paper I). Here is a follow-up study of this system with
more complete $uv$ coverage, extending to both shorter and longer $uv$
spacings, in 1.33 mm continuum, \cCO{} ($J=2-1$), and SO ($N_J=5_6-4_5$)
emission obtained with the Submillimeter Array (SMA)\footnote{The
Submillimeter Array is a joint project between the Smithsonian Astrophysical
Observatory and the Academia Sinica Institute of Astronomy and Astrophysics,
and is funded by the Smithsonian Institution and the Academia Sinica.}
\citep{Ho2004}.
This study may show for the first time a change of rotation profile in an
envelope and thus how a disk can be formed inside an envelope.

\section{Observations}

Observation toward the HH 111 system was first carried out with the
SMA on 2005 December 5 in the compact-north configuration, with the results
reported in \citet{Lee2009HH111}.
Follow-up observations were then carried out on 2010 January 24 in the extended
configuration for longer $uv$ spacings to resolve the structure of the
envelope and on 2010 March 26 in the sub-compact configuration for
shorter $uv$ spacings to restore the large-scale structure.
The 230 GHz
band receivers were used to observe the \aCO{} ($J=2-1$), \bCO{} ($J=2-1$),
\cCO{} ($J=2-1$), and SO (\SOta) lines simultaneously with the 1.33 mm continuum.
The rest frequency of the SO line is assumed to be 
219.949391 GHz \citep{Lovas1992}.
Note that the \aCO{} and \bCO{} lines trace mainly the outflow interaction and
will be presented in a future publication.

In the follow-up observations, the velocity resolution in SO line was increased to
$\sim$ 0.28 \vkm{} per channel, similar to that of \cCO{}.
Also, only one single pointing toward the central region was observed
in order to have enough sensitivity.
The visibility data were calibrated with the MIR package.
The flux uncertainty was estimated to be $\sim$ 15\%.
The calibrated visibility data were combined with that in the compact-north
configuration and then imaged with the MIRIAD package, as described in
\citet{Lee2009HH111}.
With various different weightings, the synthesized beams can have
a size from \arcsa{1}{6}$\times$\arcsa{1}{5} to
\arcsa{1}{2}$\times$\arcsa{1}{0}.
The rms noise levels are $\sim$ 0.06 \Jyb{} in the 
\cCO{} and SO channel maps with a beam of \arcsa{1}{3}$\times$\arcsa{1}{1},
and 1.6 \mJyb{} in the continuum map with a beam of \arcsa{1}{4}$\times$\arcsa{1}{1}.
The velocities in the channel maps are LSR.

\section{Results}


As in \citet{Lee2009HH111}, the results are presented in comparison to a
mosaic image based on the Hubble Space Telescope (HST) NICMOS image ([FeII]
1.64 \micron{} + \H2{} at 2.12 \micron{} + continuum) obtained by
\citet{Reipurth1999}, which shows two sources in the infrared,
reflection nebulae that trace the illuminated outflow cavity walls, and the
jet in the system. The
two sources were also detected in 3.6 cm
continuum by the VLA as the VLA 1 and 2 sources, respectively, at
$\alpha_{(2000)}=\ra{05}{51}{46}{254}$,
$\delta_{(2000)}=\dec{+02}{48}{29}{65}$ and
$\alpha_{(2000)}=\ra{05}{51}{46}{07}$,
$\delta_{(2000)}=\dec{+02}{48}{30}{76}$ \citep{Reipurth1999,Rodriguez2008}.
These VLA positions are more accurate than the NICMOS positions and 
are thus adopted here as the source positions.
The systemic velocity in this system is assumed to be 8.90$\pm$0.14 \vkm{}
LSR, as discussed in Section \ref{ssec:spec}.
Throughout this paper, the velocity is relative to this systemic value.

\subsection{1.33 mm Continuum Emission} \label{ssec:cont}

\begin{figure} [!hbp]
\centering
\includegraphics[angle=-90,scale=1]{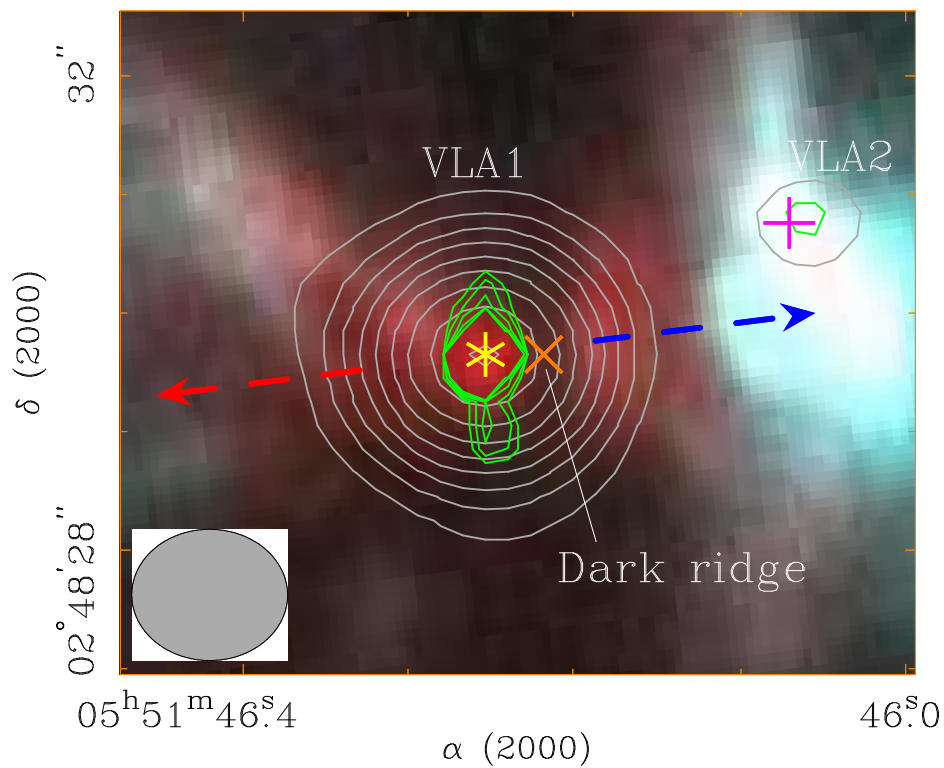}
\figcaption[]
{
The 1.33 mm continuum emission contours overplotted on the HST NICMOS
image adopted from \citet{Reipurth1999}. The gray and green contours are from
the restored map and the CLEAN component map, respectively, of the continuum
emission. The asterisk, cross, and ``x'' mark the positions of the VLA 1 and 2
sources, and the center of the dark ridge, respectively.
The gray contour levels are $5\sigma (1-r^n)/(1-r)$, where $r=1.3$,
n=1,2,3.., and $\sigma=1.6$ \mJyb{}. The green contour levels start at 4\%, 
with a step of 3\%, of the peak value.
The synthesized beam has a size of
\arcsa{1}{4}$\times$\arcsa{1}{1}. The blue and red arrows indicate the
orientations of the blueshifted and redshifted parts of the jet,
respectively.
\label{fig:cont}
}
\end{figure}

As shown in  Figure \ref{fig:cont}, the continuum emission is now
spatially resolved into two components, a bright one associated
with the VLA 1 source  
and a much fainter one ($\sim$ 5 $\sigma$ detection with 1
$\sigma=1.6$ \mJyb{}) associated with the VLA 2 source. 
The one associated
with the VLA 1 source is slightly elongated
perpendicular to the jet axis. This elongation
is better seen in the CLEAN
component map, which shows a faint disk-like structure
extending to $\sim$ \arcsa{0}{7} (280 AU) to the north and
\arcsa{0}{9} (360 AU) to the south. Note that in the CLEAN component
map, the central peak at the VLA 1 position has
a FWHM size of $\lesssim$ \arcsa{0}{6} (240 AU) and $\sim$ 90\% of the
flux, but no emission is detected toward the dark
ridge position at the center of the reflection nebulae likely
because of a lack of sensitivity.
As discussed in \citet{Lee2009HH111}, the emission is thermal emission
from dust. Thus, the emission is believed to come mainly
from a dusty disk, with the inner part of which already seen with a size of
$\sim$ \arcsa{0}{15} (60 AU) at 7 mm \citep{Rodriguez2008}. 
The total flux of this
emission is $\sim$ 300$\pm$40 mJy, slightly higher than that found
in \citet{Lee2009HH111} because of the inclusion of the shorter $uv$
spacings. In \citet{Lee2009HH111}, the dust was found to have a 
temperature of 41-64 K, with a mass opacity of 0.026 cm$^2$ g$^{-1}$
and the emission being optically thin at 1.33 mm.
Therefore, the gas and dust associated with this emission have
a mass of
$\sim 0.13 \pm 0.03$ \solarmass{}.
Most of this mass, however, is from the central unresolved peak.
On the other hand, the total flux toward the VLA 2
source is $\sim$ 11$\pm2$ mJy. The mass would be $\sim$ 0.005 \solarmass{}
if the emission has the same properties as that around the VLA 1 source.

\subsection{Line Emission}

\subsubsection{Spectra} \label{ssec:spec}

\begin{figure} [!hbp]
\centering
\includegraphics[angle=-90,scale=0.7]{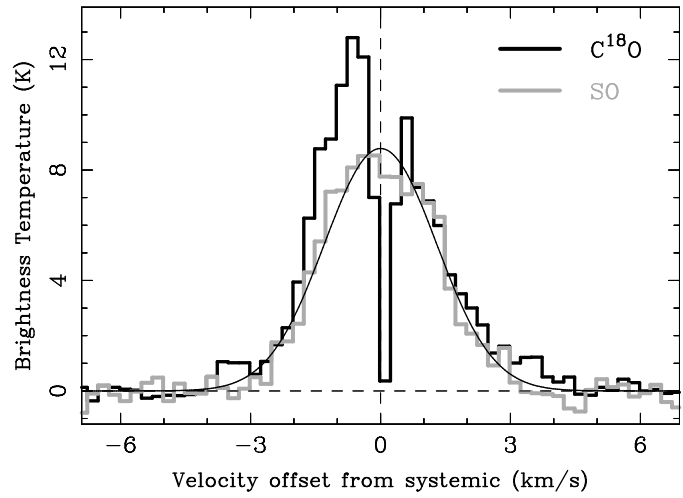}
\figcaption[]
{\cCO{} and SO spectra toward the VLA 1 source averaged over an elliptical region of 
\arcs{2}$\times$\arcs{1} in size with the major axis in the equatorial plane.
The SO spectrum can be roughly described by a Gaussian line profile (solid
curve), and its line center can be used here to define the systemic velocity.
\label{fig:spec}
}
\end{figure}

Figure \ref{fig:spec} shows the \cCO{} and SO spectra toward the VLA 1
source. These lines are detected upto $\sim$ $\pm$4.0 \vkm{} from the
systemic velocity.  The SO spectrum is rather symmetric and can be
roughly described by a Gaussian line profile, and its line center can be
used here to define the systemic velocity. Since part of the SO emission
comes from the outflow as discussed later, further observation at higher resolution is
needed to refine the systemic velocity by separating the outflow from the
envelope. The \cCO{} spectrum shows a double-peaked profile with a dip near
the systemic velocity, and with the blueshifted peak brighter than the
redshifted peak (i.e., the blue asymmetry). As discussed in
\citet{Lee2009HH111}, the dip with the blue asymmetry could be due to
self-absorption because of an infall motion in the envelope and the missing
flux resolved out by the interferometer. The SO emission does not show a
dip, likely because it is optically thin due to low SO abundance.

\subsubsection{Morphologies at different velocities}

In order to see how the envelope structure changes with
velocity, the line emission is divided into 3 velocity ranges: low (0$-$1
\vkm{}),
medium (1$-$2 \vkm{}), and high (2$-$3.7 \vkm{}) velocity ranges, on the
redshifted and blueshifted sides.

\begin{figure} [!hbp]
\centering
\includegraphics[angle=0,scale=0.60]{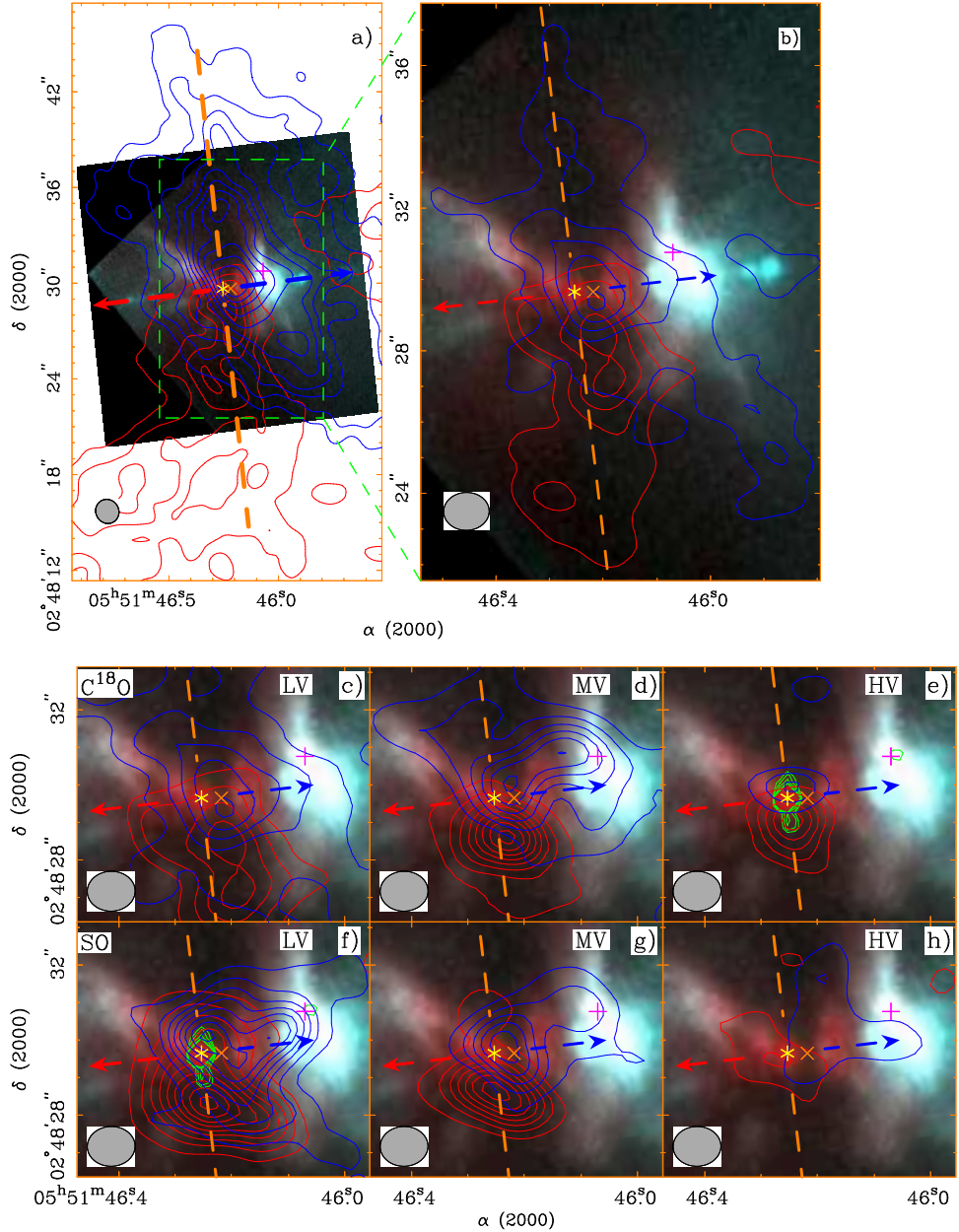}
\figcaption[]
{\small
Contours of blueshifted (blue) and redshifted (red)
emission of \cCO{} and SO on the HST NICMOS image. 
The asterisk, cross, ``x'', and the blue and red arrows
 have the same meanings as those in Figure
\ref{fig:cont}.
The dashed line indicates the equatorial plane perpendicular to the jet axis.
The beam has a size of \arcsa{1}{6}$\times$\arcsa{1}{5} in (a) and
\arcsa{1}{3}$\times$\arcsa{1}{1} for the rest. 
{In (a), a large beam is used to show the large-scale structure of the envelope.}
\tlabel{Top two panels} Low-velocity blueshifted and redshifted emission of \cCO{}.
In (a), the contour levels start at 4$\sigma$ with a step of 4$\sigma$,
where $\sigma=0.032$ \Jybk{}.
In (b), the contour levels start at 3$\sigma$ with a step of 3$\sigma$,
where $\sigma=0.03$ \Jybk{}.
\tlabel{Middle three panels} Low-, medium-, and high-velocity (i.e, LV, MV,
and HV) blueshifted and redshifted emission of \cCO{} in the central region.
The contour levels are the same as those in (b).
\tlabel{Bottom three panels} Same as those in the middle three panels, but for SO.
The green contours in \tlabel{e} and \tlabel{f} are the same as those
in Figure \ref{fig:cont}, showing the CLEAN component map of the continuum
emission.
\label{fig:linemapl}
}
\end{figure}

The structure of the envelope can be clearly seen in \cCO{}
at low velocity, as shown
in Figure \ref{fig:linemapl}a.
The envelope is
seen extending to $\sim$ \arcs{16} (6400 AU) to the north and south from
the VLA 1 source, about twice as extended as that seen in
\citet{Lee2009HH111}. It is well confined in the equatorial plane
perpendicular to the jet axis, 
fitting right in between the cavity walls outlined by the reflection
nebulae. The detailed structure, however, is complicated. In particular, the
emission in the north beyond \arcs{12} out from the source is deviated away from
the equatorial plane and the emission in the south is curvy.
With respect to the VLA 1 source, the
blueshifted emission is to the north and the redshifted one is to the south, as
expected for a rotating envelope. 
As we zoom into the inner region at higher resolution, the emission appears to be aligned with the dark ridge at $\sim$
\arcsa{0}{5} in the west of the VLA 1 source (see Figure
\ref{fig:linemapl}b and c for the central region). The blueshifted emission also 
arises from the outflow, extending to the west
along the jet axis, to the northeast along the cavity wall, and to the
southwest. At medium velocity, the emission shrinks toward the VLA 1 source
(Figure \ref{fig:linemapl}d),
as expected if the velocity of the emission increases toward the source.
The blueshifted peak is closer to the source than the redshifted peak.
In addition, both the blueshifted and redshifted peaks shift to
the east toward the equatorial plane of the
VLA 1 source, as compared to those at low velocity.
Note that the blueshifted emission also
extends toward the VLA 2 source and along the jet.
At high velocity, the emission shrinks even closer to the
source (Figure \ref{fig:linemapl}e). In addition,
the blueshifted and redshifted peaks shift
further to the east into the equatorial plane of the VLA
1 source, coincident with the disk-like structure seen 
in the continuum in the CLEAN component map.
Thus, the motion at high velocity is highly dominated by
the rotation, with the emission in the equatorial plane. Again, the blueshifted peak is closer to the source than the
redshifted peak, and here by $\sim$ \arcsa{0}{3}.

Unlike the \cCO{} emission, the SO emission is compact with a size of
$\lesssim$ \arcs{3}, tracing the innermost part of the envelope. At low velocity,
the blueshifted and redshifted peaks are seen on the opposite
sides of the VLA 1 source in the equatorial plane (Figure
\ref{fig:linemapl}f), tracing the
rotation motion. That both the redshifted and blueshifted emission also extend to
the opposite sides of the source could be due to an additional motion,
either infall or outflow.
In addition, the blueshifted emission also extends to the
VLA 2 source. At medium velocity, the emission shrinks toward the VLA 1
source (Figure \ref{fig:linemapl}g).  The peak of the blueshifted emission,
however, is shifted to the west of the VLA 1 source, likely due to the
contamination of the emission extending to the VLA 2 source and the emission
at the outflow base. At high velocity, the emission is very weak (Figure
\ref{fig:linemapl}h). Both the redshifted and blueshifted emission seem to
be contaminated by the outflow emission, extending along the cavity
walls and along the jet axis.

\subsubsection{Rotation motion}\label{ssec:rotation}

\begin{figure} [!hbp]
\centering
\includegraphics[angle=-90,scale=0.9]{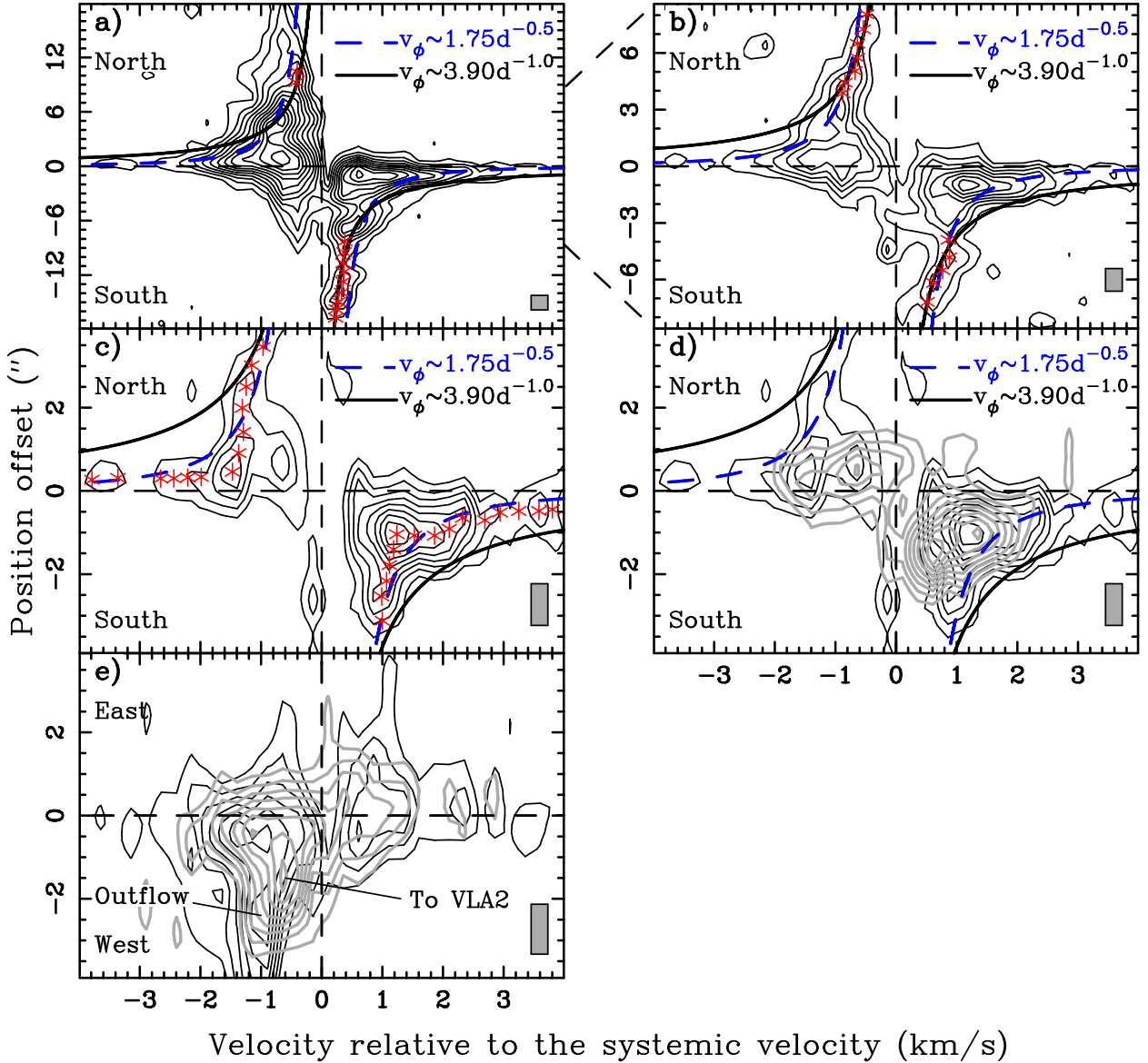}
\figcaption[]
{\small
Position-velocity (PV) diagrams in \cCO{} (black)
and SO [gray in \tlabel{d} and \tlabel{e}]
centered at the VLA 1 source.
The gray boxes in the lower-right corners show the velocity and
spatial resolutions of the PV diagrams.
\tlabel{a-d} PV diagrams 
cut perpendicular to the jet axis with increasing angular
resolution toward the center. The resolutions are
\arcsa{1}{6}$\times$\arcsa{1}{5} in \tlabel{a},
\arcsa{1}{3}$\times$\arcsa{1}{1} in \tlabel{b}, and
\arcsa{1}{2}$\times$\arcsa{1}{0} in \tlabel{c} and \tlabel{d}. 
The asterisks mark the data points that are used in Figure \ref{fig:pvdat}.
Solid curves are derived from the rotation that has constant specific
angular momentum. Dashed curves are derived from the Keplerian rotation.
Here $d=R/\Ro$, with $R$ being the radial distance from the source and $\Ro=$\arcs{1}
(400 AU).
\tlabel{e} PV diagrams cut along the jet
axis. The resolution is the same as that in \tlabel{b}. The label
``To VLA2" means that the SO emission there extends to the VLA 2 source.
For \cCO{}, the contours start at 2$\sigma$ and have a
step of 2$\sigma$, where $\sigma=$ 0.06 \Jyb{} in \tlabel{a}, 0.045 \Jyb{}
in \tlabel{b} and \tlabel{e}, and 0.037 \Jyb{} in \tlabel{c} and \tlabel{d}.
For SO, the contours start at 3$\sigma$ and have a
step of 2$\sigma$, where $\sigma=$ 0.045 \Jyb{} in \tlabel{d} and \tlabel{e}.
\label{fig:pvs}
}
\end{figure}

The rotation profile in the envelope can now be 
better constrained than that in \citet{Lee2009HH111},
with more complete $uv$ coverage.
Figures \ref{fig:pvs}a through \ref{fig:pvs}c
show the position-velocity (PV) 
diagrams centered at the VLA 1 source cut perpendicular to the jet axis in
\cCO{}, with increasing resolution toward the inner region.
The blueshifted emission is seen mainly to the north and redshifted
emission mainly to the south, confirming that the motion in the envelope is
dominated by rotation. The blueshifted emission beyond $\sim$ \arcs{12} in
the north shows a velocity increasing with the distance from the source
(Figure \ref{fig:pvs}a).
That part of the emission is seen deviated from the equatorial plane (see
Fig. \ref{fig:linemapl}a) and thus may trace an interaction with the
surrounding. Except for that part, the rotation velocity increases toward
the source and thus could be Keplerian or that 
has constant specific angular momentum as seen in some infalling
envelopes in the Class I phase \cite[see,
e.g.,][]{Ohashi1997a,Momose1998,Lee2005L1221}.

\begin{figure} [!hbp]
\centering
\includegraphics[angle=-90,scale=0.7]{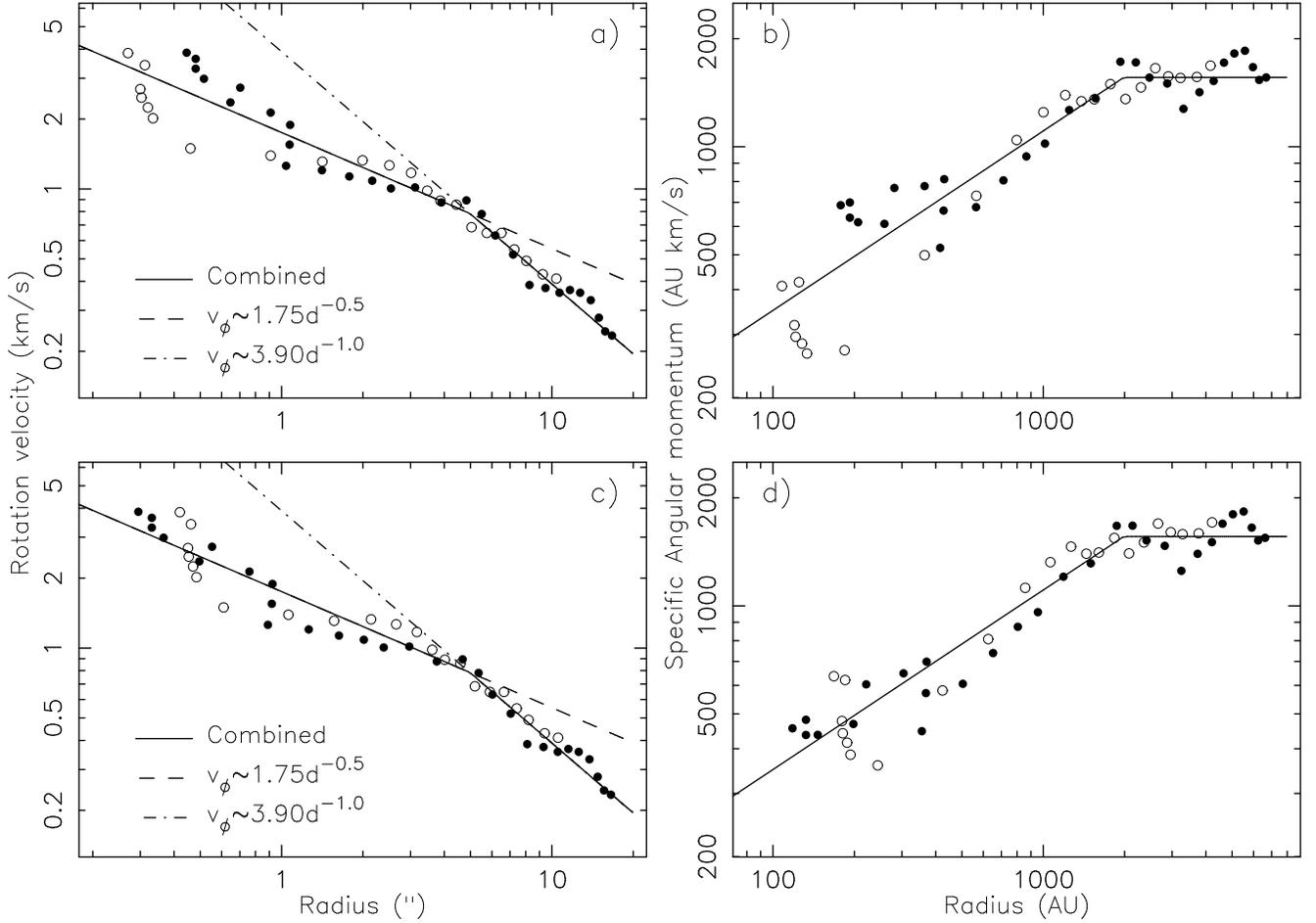}
\figcaption[]
{ Rotation velocity and implied specific angular momentum plotted as a function of
the radial distance from the source, with both corrected for the inclination angle.
Open and filled circles are from the blueshifted and redshifted emission,
respectively.
Here $d=R/\Ro$, with $R$ being the radial distance from the source and $\Ro=$\arcs{1}
(400 AU).
(c) and (d) are derived from (a) and (b),
respectively, by shifting the center of the envelope by \arcsa{0}{15} to the
south.
\label{fig:pvdat}
}
\end{figure}

Figures \ref{fig:pvdat}a and \ref{fig:pvdat}b show the rotation velocity
($v_\phi)$ and the implied specific angular momentum ($l=R v_\phi$), 
respectively, as a function of $R$, which is the radius 
measured from the source.
The data points are extracted from the PV diagrams 
with increasing resolution toward the inner region
and marked as ``x''s in Figures \ref{fig:pvs}a through \ref{fig:pvs}c.
Both the rotation velocity and specific angular momentum 
are corrected for the inclination angle assuming that the
envelope is perpendicular to the jet axis and thus has an inclination angle
of 10\degree{} to the line of sight \citep{Reipurth1992}. 
As can be seen, the rotation profile can be roughly divided into two parts. 
For $R\gtrsim$ \arcs{5} (2000 AU), the rotation velocity 
can be roughly reproduced by
the rotation that has constant specific angular momentum (see also
Figures \ref{fig:pvs}a through \ref{fig:pvs}c), 
with $v_\phi \sim 3.90 (R/\Ro)^{-1}$ \vkm{} and $\Ro=$ \arcs{1} (or 400 AU).
The specific angular momentum there is almost constant at $\sim$ 1550 AU
\vkm{}. For $R <$ \arcs{5}, the rotation velocity increases slower toward
the source and is better described by a Keplerian
law (see also Figures \ref{fig:pvs}a through
\ref{fig:pvs}c), with $v_\phi \sim 1.75 (R/\Ro)^{-0.5}$ \vkm{}. 
The specific angular momentum decreases steadily to 
$\sim$ 300$-$400 AU \vkm{} at $\sim$ 100 AU (\arcsa{0}{25}).
Note that in Figures \ref{fig:pvdat}a and \ref{fig:pvdat}b,
for $R \lesssim$ \arcs{1},
the blueshifted data points and the redshifted data points are separated from each other.
As mentioned, at high velocity, the peak of the blueshifted emission in the
north is closer
to the source by $\sim$ \arcsa{0}{3} than that of the redshifted emission in
the south (Figure
\ref{fig:linemapl}e). This difference in distance is also seen in the PV diagram in 
Figure \ref{fig:pvs}c. Thus, it is possible that the center of the envelope
is actually located at \arcsa{0}{15} to the south of the VLA 1 source.
If so, then the blueshifted data points and
the redshifted data points will move closer to each other, as seen
in Figures \ref{fig:pvdat}c and \ref{fig:pvdat}d. Observations at high
angular resolution are needed to confirm this.

Rotation is also seen in the compact SO emission near the source. As shown in 
Figure \ref{fig:pvs}d, the PV structure of the SO emission overlaps
with that of the \cCO{} emission near the source, tracing the rotation there.
As in \cCO{}, the center of
the PV structure seems to be located slightly to the south.

\subsubsection{Infall motion}

The infall motion can be studied with the PV diagrams cut along the jet axis
in \cCO{} and SO, as shown in Figure \ref{fig:pvs}e. The \cCO{} emission at
$\sim$ $-$1 \vkm{} on the blueshifted side with the position offset $<$
$-$\arcs{1} is from the outflow emission extending to the west along the jet
axis as seen in the low and medium-velocity maps (see Figures
\ref{fig:linemapl}c and d).  This blueshifted emission
in the west also extends close to the source, contaminating the envelope
emission. Also, the SO emission on the blueshifted side at $\sim$ ($-$0.6
\vkm, $-$\arcsa{1}{5}) extends to the VLA 2 source, and thus should be
excluded for studying the infall motion. Excluding those parts of the
emission, the
\cCO{} and SO PV structures appear to be similar and both are roughly symmetric
about the source, with the blueshifted emission shifted slightly to the west
and the redshifted to the east. 
With the near end of the envelope tilted to the east,
these PV structures, together with the double-peaked line profile with
a blue asymmetry and an absorption dip at the systemic velocity (Fig
\ref{fig:spec}), suggest an
infall motion in the envelope, as discussed in \citet{Lee2009HH111}.

\section{A flattened rotating envelope model}

In order to roughly derive the structure, the
physical properties (i.e., density and temperature), and the infall velocity
of the envelope, a simple flattened envelope model is used to reproduce the PV
diagrams, spectrum, and integrated intensity map of the \cCO{} emission.
In this model, the envelope has an inner radius of $\Rin$, an outer
radius of $\Rout$, and a half-opening angle of $\theta_0$, so that
its half thickness, $H$, can be given in the cylindrical
coordinates $(R,\theta,z)$ by
\begin{equation}
H(R) = \textrm{max}(H_0, R \tan \theta_0)
\end{equation}
where $\theta_0$ is measured from the equatorial plane, and
$H_0$ is the minimum value of the half thickness near the source.
The number density of molecular hydrogen in the envelope is assumed to be given by
\begin{equation} 
n (R,z) = \no \Big(\frac{\sqrt{R^2+z^2}}{\Ro}\Big)^p
\end{equation}
where $\Ro=400$ AU (\arcs{1}, as in Section \ref{ssec:rotation}),
$\no$ is the density at $\Ro$ and $p$ is a power-law
index assumed to be $-1.5$, as in many theoretical infalling models
\cite[see, e.g.,][]{Shu1977,Nakamura2000}.
The abundance of \cCO{} relative to molecular
hydrogen is assumed to be constant and given by $1.7\times10^{-7}$, as in
\citet{Lee2009HH111}.
The temperature of the envelope
is uncertain and assumed to be given by 
\begin{equation} 
T (R) = \To (\frac{R}{\Ro})^q 
\end{equation}
where $\To$ is the temperature at $\Ro$ and
$q$ is a power-law index. 

As discussed in Section \ref{ssec:rotation}, 
the rotation is assumed to change from that
with constant specific angular momentum in the outer part to that of
Keplerian in the inner part. i.e.,
\begin{eqnarray}
v_\phi(R)= \left\{ \begin{array}{cl}
v_k \big(\frac{R}{\Ro}\big)^{-0.5} & \;\;\textrm{if}\;\; R < R_t,  \\ 
v_c \big(\frac{R}{\Ro}\big)^{-1} & \;\;\textrm{if}\;\; R \geq R_t
\end{array}  \right.
\end{eqnarray}
with the transition radius
\begin{equation}
R_t = \Ro (\frac{v_c}{v_k})^2
\end{equation}
at which the Keplerian rotation is
the same as that with constant specific angular momentum.
The infall motion is uncertain. It is assumed 
to be in the radial direction and given by
\begin{equation}
v_r(R,z) = v_{r0} \Big(\frac{\sqrt{R^2+z^2}}{\Ro}\Big)^{-0.5} 
\end{equation}
as in a collapsing envelope. 

In the model, radiative transfer is used to calculate the emission, with an
assumption of local thermal equilibrium. For simplicity, the line width is
assumed to be given by the thermal line width only. The channel maps of the
emission derived from the model are used to calculate the model visibility
data with the observed $uv$ coverage. This model visibility data are used to
obtain the channel maps, and then the integrated intensity map, spectrum,
and PV diagrams to be compared with the observations.

\begin{figure} [!hbp]
\centering
\includegraphics[angle=-90,scale=0.7]{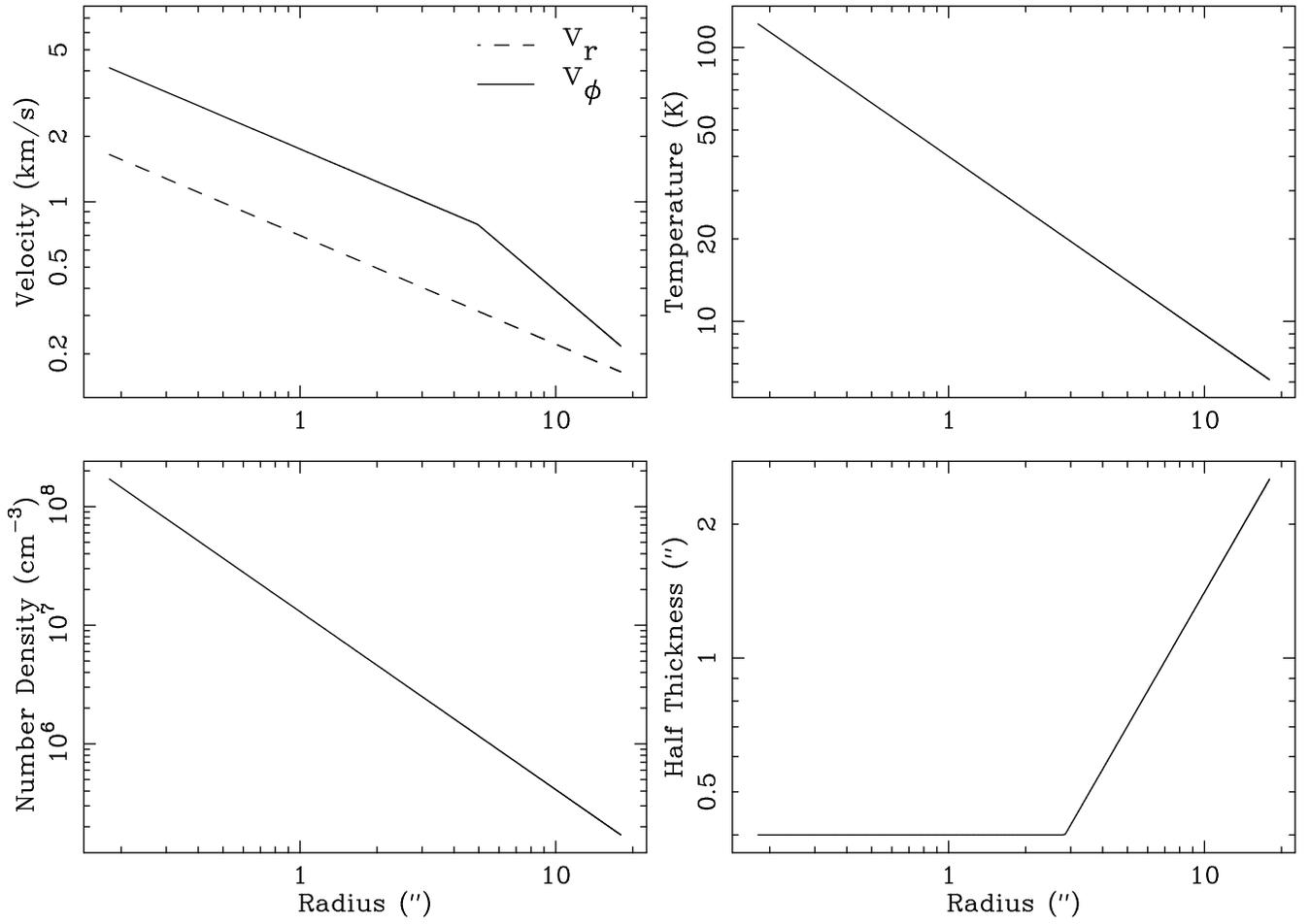}
\figcaption[]
{The best-fit parameters in our model for the \cCO{} envelope.
\label{fig:fitpar}
}
\end{figure}

\begin{figure} [!hbp]
\centering
\includegraphics[angle=-90,scale=0.7]{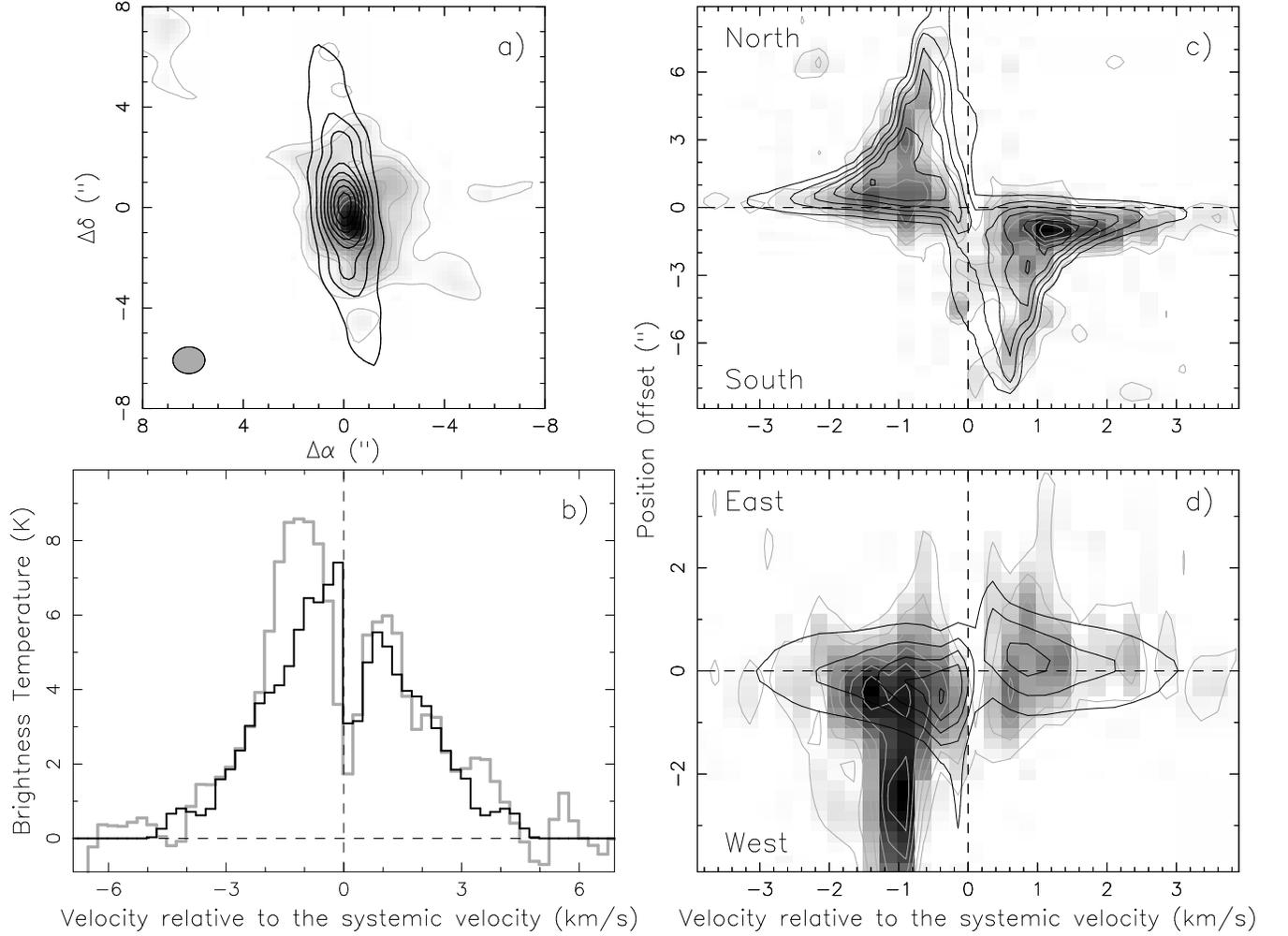}
\figcaption[]
{A simple model fitting to the integrated intensity map, spectrum, and PV diagrams in \cCO{}.
Black contours and spectrum are from the model.
Gray contours with image and spectrum are from the observations.
\tlabel{a} shows the integrated intensity maps. The contours start at 3$\sigma$ and have a
step of 3$\sigma$, where $\sigma=$ 0.06 \Jybk{}.
\tlabel{b} shows the spectra toward the VLA 1 source.
\tlabel{c} shows the PV diagrams cut
perpendicular to the jet axis, as in Figure
\ref{fig:pvs}\tlabel{b}. \tlabel{d} shows the PV diagrams cut along the jet
axis, as in Figure \ref{fig:pvs}\tlabel{e}.
\label{fig:modelC18O}
}
\end{figure}

Figure \ref{fig:modelC18O} shows the comparison of the best model with the
observations. As mentioned earlier, the envelope is assumed 
to have an inclination angle of 10\degree{} to the line of sight,
with the nearside tilted to the east. The parameters
that are used to reproduce the observations are
$v_k \sim 1.75\pm0.2$ \vkm{}, $v_c \sim 3.90\pm0.4$ \vkm, 
and $v_{r0} \sim -0.7\pm0.2$ \vkm{} for the envelope kinematics,
$\Rin \sim$ \arcsa{0}{15} (60 AU), $\Rout \sim$ \arcs{18} (7200 AU), $H_0
\sim$ \arcsa{0}{4} (160 AU), and $\theta_0 \sim$ 8$\pm2$\degree{} for the envelope
structure, and
$n_0 \sim 1.3\pm0.3\times10^7$ \cmc{}, $T_0\sim 40$ K, and $q \sim
-0.65\pm0.1$ for the envelope physical
properties (see also Figure \ref{fig:fitpar} for the parameter profiles).
The values of $v_k$ and $v_c$ are set to those found in section
\ref{ssec:rotation}, resulting in the transition radius of
$R_t \sim$ \arcs{5} (2000 AU).
The mass that supports the Keplerian rotation is $\sim$
1.38 \solarmass{}. Thus, the mass of the central source, which can be estimated by
subtracting from this mass the mass of the dusty disk close
to the source, is $\sim$ 1.25 \solarmass{}.
Here, the inner radius $\Rin$ is set to where it has a Keplerian rotation velocity
of $\sim$ 4.5 \vkm{}.
$T_0$ is set to be the lower end of the dust temperature \citep{Lee2009HH111}.
As can be seen, the simple model can reproduce the observed PV diagrams,
spectrum, and integrated intensity map reasonably well. In the model, the intensity
near the systemic velocity is higher than that in the observations, probably
because a large-scale envelope (with a size larger than \arcs{20}) or core is
needed to resolve out the emission there. Also, the blueshifted emission
around $-$1 \vkm{} is less than that in the observations because the
observed blueshifted emission there is contaminated by the outflow emission
(Figure \ref{fig:modelC18O}b).
Moreover, as compared to the observations, the PV structure perpendicular to the
jet axis is shifted slightly to the north (Figure
\ref{fig:modelC18O}c), probably because the
center of the envelope is indeed located slightly to the south, as discussed
earlier.

As a result, the \cCO{} envelope can be considered as a flattened envelope
that extends from $\sim$ 60 to 7000 AU out from
the source, with a small opening angle of $\sim$ 16\degree{}.
It has a differential rotation, with the outer part ($\sim$ 2000--7000 AU)
better
described by a rotation that has constant specific angular momentum and the
inner part ($\sim$ 60--2000 AU) by a Keplerian rotation. The transition
radius between the two parts is $\sim$ 2000 AU.
The mass of the envelope (including helium and molecular hydrogen) 
is given by the following integration:
\begin{eqnarray}
M_e &=& 2.8 \mH \no \int_{\Rin}^{R} \int_{-H}^{H} 
 (\frac{\sqrt{R^2+z^2}}{\Ro})^{-1.5} \;2 \pi R dR dz \nonumber
\end{eqnarray} 
Thus, the mass of the whole envelope extending from 60 to 7000 AU is $\sim$
0.62 \solarmass{}, about 45\% of the mass of the central source. On the
other hand, the mass of the inner part of the envelope extending from 60 to
2000 AU is $\sim$ 0.13 \solarmass{}. 
Since this mass is only $\sim$ 10\% of the mass of the
central source, the rotation in the inner part of the envelope can indeed
become Keplerian.  The envelope seems to have a small infall motion.
The amount of the infall
motion, however, could be overestimated due to the outflow contamination in the
unresolved PV structure along the jet axis. Observation at higher
resolution is really needed to confirm the infall motion by separating the
outflow from the envelope.

\section{Discussion}

\subsection{Comparing to the previous results}

With a more complete $uv$ coverage, the envelope is detected to a two
times larger extent at upto three times higher resolution than that in
\citet{Lee2009HH111}. The infall velocity can now be better determined with
less outflow contamination, and is found to be $\sim$ 40\% of that derived
in \citet{Lee2009HH111}. The rotation profile in the inner part of the
envelope can now be better constrained with a larger extent of the envelope,
and is found to be more like Keplerian but with a 30\% higher velocity than
that found in \citet{Lee2009HH111}. Thus, the inner part of the envelope is
no longer to be consistent with a dynamically infalling disklike envelope
(or pseudo-disk), as
suggested in \citet{Lee2009HH111}. 
Previously, the mass of the central source was derived from the
infall velocity assuming a dynamical infall and found to be $\sim$ 0.8 \solarmass{},
but now is from the Keplerian rotation velocity and found to be $\sim$ 1.25
\solarmass{}. In addition, the compact SO envelope near the
source is now better resolved and thus better separated from the outflow
cavity walls. The PV structure that was seen at the outer edge of the SO
envelope, with the velocity decreasing with the decreasing distance from the
source, is not seen here anymore. That PV structure is expected to be from the
outflow walls \citep{Lee2000}. Thus,
the outer edge of the SO envelope does not trace the inner edge of a
pseudo-disk, as suggested in \citet{Lee2009HH111}. As a result, the previous
comparison of the inner part of the envelope to a collapsing magnetized rotating toroid of
\citet{Allen2003} and \citet{Mellon2008} is not appropriate anymore.

\subsection{Collapsing envelope}

The outer part of the envelope seems infalling toward the center
with constant specific angular momentum, as found in other Class I
sources \citep{Ohashi1997a,Momose1998,Lee2005L1221}.
The kinematics there is qualitatively consistent with that predicted in some
infalling disklike envelopes (or pseudo-disks), either magnetized
\cite[][]{Terebey1984,Krasnopolsky2002}
 or nonmagnetized \cite[see, e.g.,][]{Nakamura2000}, where the material is
infalling toward the center with its angular momentum. The
infall velocity is $\sim$ 80\% of the rotation velocity at $\sim$ 7000 AU at
the edge of the envelope, and it drops to 40\% of the rotation velocity at
$\sim$ 2000 AU at the transition radius. Therefore, in details, this part of
the envelope is actually different from those collapsing envelopes found in
other Class I sources
\citep{Ohashi1997a,Momose1998,Lee2005L1221}, where the infall velocity is larger than
the rotation velocity.
The HH 111 is probably in a later stage than those
sources, with the motion already dominated by the rotation.
The material in the envelope here
is slowly spiraling inward and the centrifugal force progressively
increases, so that the rotation can indeed become
Keplerian at $\sim$ 2000 AU.
The infall rate in this part of the envelope is constant and given by
\begin{eqnarray}
\dot{M} =   2.8 \mH \int n v_r r^2 d\Omega 
\approx  4.2\times 10^{-6}\;{M_\odot}\; \textrm{yr}^{-1} \nonumber
\end{eqnarray}
This infall rate is similar to that seen in other Class I sources
\citep{Ohashi1997a,Momose1998}. Note that the infall rate here is a
factor of 2 lower than that derived in \citet{Lee2009HH111} because of a
factor of 2 lower in the infall velocity derived here.

An accretion shock is expected to occur at the transition radius, see e.g.,
\citet{Yorke1999}. This accretion shock has also been suggested to be
present in the younger protostellar system L1157 \citep{Velusamy2002} at 
a much smaller radius at $\sim$ 400 AU at a much younger age in the Class 0 phase.
Here in HH 111, the isothermal sound speed in the envelope, which can be
derived from the temperature, is $\sim$ 0.15 \vkm{} at $\sim$ 7000 AU and
0.22 \vkm{} at 2000 AU. Thus, in the outer part of the envelope, the
infall velocity is
$\sim$ 1.1$-$1.4 times the isothermal sound speed, and is thus only slightly
supersonic at that large distance. As a result, unlike that in L1157, the
accretion shock is expected to be weak at the transition radius
and thus not easy to be detected.

\subsection{2000 AU Sub-Keplerian disk?} \label{ssec:skdisk}

In the inner part of the envelope, where the rotation has become
Keplerian, the infall velocity is uncertain and assumed to follow
the same velocity profile as that in the outer part. It is
$\sim$ 40\% of the rotation velocity. It increases from 1.4 to 2 times the
isothermal sound speed from the transition radius to the vicinity of the
source at $\sim$ 240 AU (\arcsa{0}{6}). However, as mentioned earlier, the
infall velocity could be overestimated and it could actually be subsonic as
expected for a rotationally-supported disk with a Keplerian rotation.
Therefore, the transition radius could indeed be the centrifugal radius,
where the centrifugal force is balanced by the gravitational force. 
If that is the case, then the rotationally supported disk is quite large
with a radius of $\sim$ 2000 AU.  A disk with such large radius
has been seen in the simulations of
\citet{Yorke1999}, as a result of a gravitational collapse of a
rotating spherical cloud. In their simulations, the angular momentum is
transported outward via tidally induced gravitational torque. The HH 111
system can be compared to their model H, in which the cloud was set up to
have a mass of 2 \solarmass{} and a specific angular momentum of $\sim$ 1333
AU \vkm{} at the cloud edge. In that model, the cloud initially has a radius
of 13333 AU and it collapses to form a quasi-static Keplerian disk
surrounded by an infalling envelope. At the age of $\sim 4\times10^5$ yr,
the disk has grown up to $\sim$ 1800 AU in radius with a central mass of
1.121 \solarmass{} (see their Figures 2), similar to the HH 111 system.
Further observations at higher resolution are needed to check if the infall
velocity here can really be subsonic.

Alternatively, this inner part of the envelope can be considered as a 2000 AU sub-Keplerian
disk with a small infall motion. Assuming that both the infall and rotation
motions are due to the gravitational force from the center, then
the mass of
the center source would be $1+0.5(v_r/v_k)^2=1.08$ times that derived
earlier, and the disk would be
rotating at $1/\sqrt{1.08} \sim$ 96\% of the Keplerian velocity.
A similar sub-Keplerian disk with a radius of $\sim$ 2000 AU is also seen in
the Class I source L1489 IRS, with a similar mass ($\sim$ 1.3
\solarmass{}) of the central source and
a similar ratio ($\sim$ 0.4) of the infall velocity to the rotation velocity
\citep{Brinch2007}. A smaller rotating disk with a radius of $\sim$ 600 AU
is also seen in the Class I source IRAS 04365+2535,
with a smaller mass of $\sim$ 0.5 \solarmass{} for the central source
\citep{Ohashi1997b}. This disk seems to have some small infall motion too 
\citep{Ohashi1997b} and thus can also be considered as a sub-Keplerian disk.
Note that the sizes of these sub-Keplerian disks seem to scale roughly linearly
with the mass of the central source.
These sub-Keplerian disks are expected to become
fully rotationally-supported Keplerian disks closer to 
the source, where the infall
velocity is expected to drop below the sound speed and thus become much
smaller than the rotation velocity.
Here in HH 111, the rotationally-supported 
disk must have already formed around the source to launch the jet.

The accretion rate in the disk can be assumed to be the same as the infall
rate in the outer part of the envelope.
 From CO bullets along the jet axis,
the mass-loss rate was estimated to be $\sim$ 4$\times 10^{-7}$ \smassrate{}
in one side of the jet \citep{Cernicharo1996} and thus $\dot{M}_j \sim$ 8$\times
10^{-7}$ \smassrate{} in two sides. Therefore, the mass-loss rate to
accretion rate ratio is $\sim$ 0.19, comparable to that predicted in
current MHD models
\citep{Shu2000,Pudritz2007}. With a velocity of $\sim$ 300-500 \vkm{}
\citep{Cernicharo1996}, the jet has a mechanical luminosity of 
$\frac{1}{2} \dot{M}_j v_j^2 \sim$ 6-16
\solarlum{}, about 30-80\% of the bolometric luminosity of the source,
which is 
$\sim$ 20 \solarlum{}
\cite[][corrected for the new distance of 400 pc]{Reipurth1989}.
Assuming a stellar mass of $M_\ast \sim 1.3$ \solarmass{} and a stellar
radius of $R_\ast \sim 3 R_\odot$,
the accretion luminosity is $L_\textrm{\small acc} = G
M_\ast \dot{M}_\textrm{\small acc}/R_\ast \sim 58$ \solarlum{}.
This luminosity is comparable to the sum of the mechanical
luminosity of the jet and bolometric luminosity of the source, as expected
if both luminosities are derived from gravitational
potential energy released in the accretion.

\subsection{SO shocks in the disk?}


The SO emission is seen in the equatorial plane, with
a mean radius of $\sim$ 400 AU (\arcs{1}). It is seen with a rotation similar to
that seen in \cCO{}. Since SO emission has been found to trace
warm and shocked regions
\citep{Codella2005,Lee2006,Lee2010HH211}, could the SO emission here trace a shock
in and around the disk? 
The peak intensity of the SO emission is $\sim 2$ \Jybk{} in a beam of
\arcsa{1}{3}$\times$\arcsa{1}{1}. Assuming that the emission is optically
thin and is from a region in LTE at a temperature of 100 K
\citep{Codella2005}, then the column density of SO is $\sim 8 \times 10^{14}$
\cms{}. Since the emission is not well resolved, this value should be
considered as a lower limit. Dividing this column density by that of H$_2$
in the model at \arcs{1}, which is $\sim$ 6.2$\times10^{22}$ \cms{}, the SO
abundance is found to be $\sim 1.3 \times 10^{-8}$, higher than that found in
the ambient material, which is $\sim$ 1-4$\times10^{-9}$
\citep{Codella2005}, but lower than that found in the jet, which is $\sim$ 2
$\times 10^{-6}$ \citep{Lee2010HH211}.
Therefore, the SO emission here
could trace a weak shock in the inner part of the disk, such as the
inner accretion shock on the disk surface
\citep{Yorke1999}. However, it is also possible that 
the SO emission traces the accretion shock between the sub-Keplerian disk and the
rotationally-supported disk.

\subsection{Ejection?}


It has been proposed that the VLA 1 and 2 sources were first formed at
the dark ridge position and then ejected to the east and west, respectively
\citep{Reipurth1999}. In our observations, the peaks of the
\cCO{} emission shift from the dark ridge position to the east to the VLA 1 position
from the low to high velocity. This shift of the emission peaks seems to
support that the VLA 1 source was indeed formed at the dark ridge position
and then ejected to the east, as proposed. The ejection was estimated to
take place only $\sim$ 3000 yrs ago \citep{Reipurth1999}. Therefore,
the outer part of the envelope that is seen at low velocity does not have
enough time to react to this ejection. It is aligned with the dark ridge
likely because it is the one that produces the dark ridge with dust extinction.
No dust emission is
detected toward the dark ridge position because of the low density in the outer
part of the envelope.
On the other hand, the inner part of the envelope that is seen at high velocity and the
disk have reacted to this ejection, carried away by the VLA 1 source, as
seen in the observations.

\section{Conclusion}

The envelope and disks in the HH 111 protostellar system can be better
studied with more complete $uv$ coverage than that in \citet{Lee2009HH111}.
The 1.33 mm continuum emission is now spatially resolved into two
components, a bright one associated with the VLA 1 source and a much fainter
one associated with the VLA 2 source, likely tracing the dusty putative
disks around the two sources. The emission around the VLA 1 source is
elongated in the equatorial plane with a radius of $\sim$ 300 AU. The
envelope is well seen in
\cCO{}, extending from $\sim$ 60 to 7000 AU out from the VLA 1 source, with
the innermost part overlapping with the dusty disk. It has a differential
rotation, with the outer part ($\sim$ 2000--7000 AU) better described by a
rotation that has constant specific angular momentum and the inner part
($\sim$ 60--2000 AU) by a Keplerian rotation. The envelope seems to have
some infall motion too. The amount of
infall motion, however, could be overestimated due to outflow contamination.
The infall velocity to rotation velocity ratio is
found to decrease from $\sim$ 0.8 at the edge of the envelope to 0.4 at the
transition radius and in the inner part of the envelope. 
Thus, the material in the outer part of the envelope seems to be slowly
spiraling inward with {its angular momentum} and the
rotation can indeed become Keplerian in the inner part. A
compact SO emission is seen around the VLA 1 source with a radius of $\sim$
400 AU and it may trace a shock such as an (inner) accretion shock around
the disk. It seems that the VLA 1 source was first formed in the dark ridge
and then ejected to the east to its current position, as previously
proposed.

\acknowledgements

I thank the SMA staff for their efforts in running and maintaining the
array. I also thank Sheng-Yuan Liu, Mike Cai, Ruben Krasnopolsky, and
Yao-Yuan Mao for fruitful conversations. I also thank the referee for the
valuable comments.

\clearpage

\end{document}